\documentclass[aps,prl,twocolumn,superscriptaddress,showpacs,preprintnumbers,amsmath,amssymb]{revtex4-1}

\usepackage{graphicx} 
\usepackage{dcolumn}  

\graphicspath{{ps}}

\begin{document}

\title{ \quad\\[1.0cm]
{\bf {\boldmath {\large Search for a dark vector gauge boson 
decaying to $\pi^+ \pi^-$ using $\eta \rightarrow \pi^+\pi^- \gamma$ decays}}}
}
\noaffiliation
\affiliation{University of the Basque Country UPV/EHU, 48080 Bilbao}
\affiliation{Beihang University, Beijing 100191}
\affiliation{Budker Institute of Nuclear Physics SB RAS, Novosibirsk 630090}
\affiliation{Faculty of Mathematics and Physics, Charles University, 121 16 Prague}
\affiliation{Chonnam National University, Kwangju 660-701}
\affiliation{University of Cincinnati, Cincinnati, Ohio 45221}
\affiliation{Deutsches Elektronen--Synchrotron, 22607 Hamburg}
\affiliation{University of Florida, Gainesville, Florida 32611}
\affiliation{Justus-Liebig-Universit\"at Gie\ss{}en, 35392 Gie\ss{}en}
\affiliation{SOKENDAI (The Graduate University for Advanced Studies), Hayama 240-0193}
\affiliation{Hanyang University, Seoul 133-791}
\affiliation{University of Hawaii, Honolulu, Hawaii 96822}
\affiliation{High Energy Accelerator Research Organization (KEK), Tsukuba 305-0801}
\affiliation{J-PARC Branch, KEK Theory Center, High Energy Accelerator Research Organization (KEK), Tsukuba 305-0801}
\affiliation{IKERBASQUE, Basque Foundation for Science, 48013 Bilbao}
\affiliation{Indian Institute of Technology Bhubaneswar, Satya Nagar 751007}
\affiliation{Indian Institute of Technology Guwahati, Assam 781039}
\affiliation{Indian Institute of Technology Madras, Chennai 600036}
\affiliation{Institute of High Energy Physics, Chinese Academy of Sciences, Beijing 100049}
\affiliation{Institute of High Energy Physics, Vienna 1050}
\affiliation{INFN - Sezione di Torino, 10125 Torino}
\affiliation{J. Stefan Institute, 1000 Ljubljana}
\affiliation{Kanagawa University, Yokohama 221-8686}
\affiliation{Institut f\"ur Experimentelle Kernphysik, Karlsruher Institut f\"ur Technologie, 76131 Karlsruhe}
\affiliation{Kennesaw State University, Kennesaw, Georgia 30144}
\affiliation{King Abdulaziz City for Science and Technology, Riyadh 11442}
\affiliation{Department of Physics, Faculty of Science, King Abdulaziz University, Jeddah 21589}
\affiliation{Korea Institute of Science and Technology Information, Daejeon 305-806}
\affiliation{Korea University, Seoul 02841}
\affiliation{Kyungpook National University, Daegu 702-701}
\affiliation{\'Ecole Polytechnique F\'ed\'erale de Lausanne (EPFL), Lausanne 1015}
\affiliation{P.N. Lebedev Physical Institute of the Russian Academy of Sciences, Moscow 119991}
\affiliation{Faculty of Mathematics and Physics, University of Ljubljana, 1000 Ljubljana}
\affiliation{Ludwig Maximilians University, 80539 Munich}
\affiliation{Luther College, Decorah, Iowa 52101}
\affiliation{University of Maribor, 2000 Maribor}
\affiliation{Max-Planck-Institut f\"ur Physik, 80805 M\"unchen}
\affiliation{School of Physics, University of Melbourne, Victoria 3010}
\affiliation{University of Miyazaki, Miyazaki 889-2192}
\affiliation{Moscow Physical Engineering Institute, Moscow 115409}
\affiliation{Moscow Institute of Physics and Technology, Moscow Region 141700}
\affiliation{Graduate School of Science, Nagoya University, Nagoya 464-8602}
\affiliation{Kobayashi-Maskawa Institute, Nagoya University, Nagoya 464-8602}
\affiliation{Nara Women's University, Nara 630-8506}
\affiliation{National Central University, Chung-li 32054}
\affiliation{National United University, Miao Li 36003}
\affiliation{Department of Physics, National Taiwan University, Taipei 10617}
\affiliation{H. Niewodniczanski Institute of Nuclear Physics, Krakow 31-342}
\affiliation{Niigata University, Niigata 950-2181}
\affiliation{Novosibirsk State University, Novosibirsk 630090}
\affiliation{Osaka City University, Osaka 558-8585}
\affiliation{Pacific Northwest National Laboratory, Richland, Washington 99352}
\affiliation{University of Pittsburgh, Pittsburgh, Pennsylvania 15260}
\affiliation{Theoretical Research Division, Nishina Center, RIKEN, Saitama 351-0198}
\affiliation{University of Science and Technology of China, Hefei 230026}
\affiliation{Showa Pharmaceutical University, Tokyo 194-8543}
\affiliation{Soongsil University, Seoul 156-743}
\affiliation{Stefan Meyer Institute for Subatomic Physics, Vienna 1090}
\affiliation{Sungkyunkwan University, Suwon 440-746}
\affiliation{School of Physics, University of Sydney, New South Wales 2006}
\affiliation{Department of Physics, Faculty of Science, University of Tabuk, Tabuk 71451}
\affiliation{Tata Institute of Fundamental Research, Mumbai 400005}
\affiliation{Excellence Cluster Universe, Technische Universit\"at M\"unchen, 85748 Garching}
\affiliation{Department of Physics, Technische Universit\"at M\"unchen, 85748 Garching}
\affiliation{Toho University, Funabashi 274-8510}
\affiliation{Department of Physics, Tohoku University, Sendai 980-8578}
\affiliation{Earthquake Research Institute, University of Tokyo, Tokyo 113-0032}
\affiliation{Department of Physics, University of Tokyo, Tokyo 113-0033}
\affiliation{Tokyo Institute of Technology, Tokyo 152-8550}
\affiliation{Tokyo Metropolitan University, Tokyo 192-0397}
\affiliation{University of Torino, 10124 Torino}
\affiliation{Virginia Polytechnic Institute and State University, Blacksburg, Virginia 24061}
\affiliation{Wayne State University, Detroit, Michigan 48202}
\affiliation{Yamagata University, Yamagata 990-8560}
\affiliation{Yonsei University, Seoul 120-749}
  \author{E.~Won}\email[Corresponding author:~~]{eunil@hep.korea.ac.kr}\affiliation{Korea University, Seoul 02841} 
  \author{I.~Adachi}\affiliation{High Energy Accelerator Research Organization (KEK), Tsukuba 305-0801}\affiliation{SOKENDAI (The Graduate University for Advanced Studies), Hayama 240-0193} 
  \author{H.~Aihara}\affiliation{Department of Physics, University of Tokyo, Tokyo 113-0033} 
  \author{S.~Al~Said}\affiliation{Department of Physics, Faculty of Science, University of Tabuk, Tabuk 71451}\affiliation{Department of Physics, Faculty of Science, King Abdulaziz University, Jeddah 21589} 
  \author{D.~M.~Asner}\affiliation{Pacific Northwest National Laboratory, Richland, Washington 99352} 
  \author{T.~Aushev}\affiliation{Moscow Institute of Physics and Technology, Moscow Region 141700} 
  \author{R.~Ayad}\affiliation{Department of Physics, Faculty of Science, University of Tabuk, Tabuk 71451} 
  \author{I.~Badhrees}\affiliation{Department of Physics, Faculty of Science, University of Tabuk, Tabuk 71451}\affiliation{King Abdulaziz City for Science and Technology, Riyadh 11442} 
  \author{A.~M.~Bakich}\affiliation{School of Physics, University of Sydney, New South Wales 2006} 
  \author{V.~Bansal}\affiliation{Pacific Northwest National Laboratory, Richland, Washington 99352} 
% ...
\collaboration{The Belle Collaboration}

\begin{abstract}
We report a search for a dark vector gauge boson 
$U^\prime$ that couples to quarks
in the decay chain $D^{*+} \to D^0 \pi^+, D^0 \to K^0_S \eta, \eta \to 
U^\prime \gamma$, 
$U^\prime \to \pi^+ \pi^-$.
No signal is found and we set a mass-dependent limit on the baryonic 
fine structure constant of $10^{-3} - 10^{-2}$ in 
the $U^\prime$ mass range of 
290 to 520 MeV/$c^2$.  This analysis is based on a 
data sample of 976 fb$^{-1}$ collected by the Belle experiment at the
KEKB asymmetric-energy $e^+e^-$ collider. 
\end{abstract}

\pacs{14.80.-j, 13.66.Hk, 14.40.Be}

\maketitle

\tighten

{\renewcommand{\thefootnote}{\fnsymbol{footnote}}}
\setcounter{footnote}{0}

 The Standard Model (SM) of particle physics cannot 
explain the nature of dark matter that is understood to have
mostly gravitational effects on visible matter, on radiation, and on the 
large-scale structure of the 
universe~\cite{ref:zwicky, ref:clowe, ref:komatsu, ref:blumenthal}. 
The dark matter can be naturally explained by
the introduction of a weakly interacting particle predicted in the
supersymmetric extension of the SM~\cite{ref:susy}. 
The absence of observation of any supersymmetric particles in hadron collider
experiments~\cite{ref:lhcsusy} motivates studies of 
new classes of models, commonly referred to as dark models, which
introduce new gauge symmetries~\cite{ref:dark_photon} and predict the existence of
new particles that couple weakly to SM particles.
Most accelerator-based experiments have focused on the dark
photon or dark particles coupling to the SM photons
\cite{ref:belledp}, though many dark models suggest 
a new gauge boson that could couple predominantly to quarks~\cite{ref:bboson, ref:tulin}.
This new dark boson  
(hereinafter referred to as the $U^\prime$ boson,
instead of $B$ as is originally proposed in 
Ref.~\cite{ref:bboson}, to avoid confusion with the SM $B$ meson)
can be produced from light 
SM meson decays through 
$P \rightarrow U^\prime \gamma$ or
$V \rightarrow U^\prime P$,
where $P$ refers to a pseudoscalar meson (e.g., $\pi^0, \eta, \eta^\prime$)
and $V$ to a vector meson (e.g., $\omega, \phi$).
Two recent experimental
limits on searches for a dark photon $A^\prime$ via $\pi^0 \to A^\prime
\gamma,  A^\prime \to e^+ e^-$~\cite{ref:wasa} 
and $\phi \to A^\prime \gamma, A^\prime \to e^+ e^-$~\cite{ref:kloe} 
can be applied to the $U^\prime$ boson search in a model-dependent way to
constrain the baryonic fine structure constant 
$\alpha_{U^\prime} \equiv g^2_{U^\prime}/(4\pi)$,
where $g_{U^\prime}$ is the universal gauge coupling between the $U^\prime$
boson and the quarks~\cite{ref:tulin}. There are also limits from 
$\eta \rightarrow \pi^0 \gamma \gamma$ and $\phi \rightarrow \eta \pi^0 \gamma$ 
decays based on their total rate, as well as 
from the analysis of hadronic $\Upsilon (1S)$ decays~\cite{ref:tulin}.

We search for $U^\prime$ bosons decaying to $\pi^+ \pi^-$ pairs using 
$\eta \rightarrow \pi^+ \pi^- \gamma$ decays,
where $\eta$ is produced in 
the decay chain $D^{*+} \to D^0 \pi^+, D^0 \to K^0_S \eta$~\cite{ref:sign}.
The kinematics here allows us 
to suppress the combinatorial background significantly.
The decay  $U^\prime \rightarrow \pi^+ \pi^-$ 
is expected to have a relatively small branching fraction 
of 2-4\%~\cite{ref:tulin}
but nevertheless provides a very clean
signature for a possible dark vector gauge boson.
The dominant decay modes are $\pi^0 \gamma$ at low $U^\prime$ mass
and $\pi^+\pi^-\pi^0$ at higher $U^\prime$ mass, however they suffer
from higher combinatorial background  
and therefore are not used in the analysis.
We use the decay $\eta \rightarrow \pi^+ \pi^- \pi^0$
to validate our event reconstruction by
measuring the branching fraction
of $\eta \rightarrow \pi^+ \pi^- \gamma$ relative to that of 
$\eta \rightarrow \pi^+ \pi^- \pi^0$.


 The data used in this analysis were recorded at 
the $\Upsilon(nS)$ resonances ($n=1,\hdots,5$) 
and 60 MeV below the $\Upsilon(4S)$ resonance with the 
Belle detector~\cite{ref:belle} at the $e^+ e^-$ asymmetric-energy
collider KEKB~\cite{ref:kekb}. The sample corresponds
to an integrated luminosity of 976 fb$^{-1}$.
 We generated two million Monte Carlo (MC) events~\cite{ref:MC} each for 
$\eta\to\pi^+ \pi^-\gamma$, $\eta\to \pi^+ \pi^-\pi^0$, and
$\eta\to U^\prime \gamma\to \pi^+\pi^-\gamma$ 
at a particular $U^\prime$ mass selected in the range from 280 to
540~MeV/$c^2$ in steps of 10~MeV/$c^2$ (\textit{i.e.,} 58 million
events in all). 
The lifetime of the $U^\prime$
is assumed to be negligible.
The $U^\prime$ samples are used to
determine the $M(\pi^+\pi^-)$ resolution. The $U^\prime$ signal shape
parameters for intermediate $U^\prime$ mass values are determined using
spline interpolation.

Except for tracks from $K^0_S$ decays, 
we require that the charged tracks originate
from the vicinity of the interaction point (IP)
with impact parameters along the
beam direction ($z$ axis) and perpendicular to
it of less than 4 cm and 2 cm, respectively. 
All such charged tracks are required to have at least
two associated hits in the silicon vertex detector (SVD),
both in the $z$ and perpendicular directions. 
Such charged tracks are identified as pions
or kaons by requiring that the ratio of particle 
identification likelihoods, $\mathcal{L}_K/
(\mathcal{L}_K+\mathcal{L}_\pi)$, constructed
using information from the central drift 
chamber (CDC), time-of-flight scintillation counters,
and aerogel threshold Cherenkov counters,
be larger or smaller than 0.6, respectively.
For both kaons and pions, the efficiencies
and misidentification probabilities are
86\% and 14\%, respectively.

For photon selection, we require the 
energy of the candidate photon to be greater than
60 MeV (100 MeV) when the candidate photon is 
reconstructed in the barrel (endcap) calorimeter
that covers $32^\circ < \theta < 130^\circ$
( $12^\circ < \theta < 32^\circ$ or 
$130^\circ < \theta < 157^\circ$)
in the polar angle $\theta$ with respect to the $+z$ axis.
To reject neutral hadrons, the ratio of the energy deposited by a photon candidate 
in the $3\times 3$ and $5 \times 5$
calorimeter arrays centered on the crystal with the largest 
signal is required to exceed 0.85.

Candidate $\pi^0$ mesons are reconstructed from pairs of $\gamma$ 
candidates; we require $M_{\gamma \gamma}$ $\in [120,150]$ MeV$/c^2$
and refit $\gamma$ momenta with the $\pi^0$ mass constraint. 

Candidate $K^0_S \to \pi^+ \pi^-$ mesons are 
reconstructed from two tracks, assumed to be pions,
using a neural network (NN) technique~\cite{Feindt:2006pm}
that uses
the following information: the $K^0_S$ momentum in the laboratory frame;
the distance along $z$ between the two track helices at their closest approach;
the $K_S^0$
flight length in the transverse plane;
the angle between the $K^0_S$ momentum and the vector
joining the $K^0_S$ decay vertex to the IP;
the angles between the
pion momenta and the laboratory-frame
direction  in
the $K^0_S$ rest frame;
the distances of closest approach in the transverse plane between the IP and
the two pion helices;
and
the pion hit information in the SVD and CDC.
We also require that the $\pi^+ \pi^-$ invariant mass be
within $\pm 9$~MeV/$c^2$ (about 3$\sigma$ in resolution~\cite{ref:wonbr})  
of the nominal $K^0_S$ mass~\cite{PDG}.

For the $\eta \rightarrow \pi^+\pi^- \gamma$ candidates, we require 
that the 
photon not be associated with a $\pi^0$ candidate and its 
transverse momentum be greater than 200 MeV/$c$ to 
remove $D^{*+} \rightarrow D^+ (\rightarrow K^0_S \pi^+ \pi^- \pi^+) \gamma$ 
background. 
For both $\eta \to \pi^+ \pi^- \gamma$
and $\eta \to \pi^+ \pi^- \pi^0$ candidates,
we perform a vertex fit with the two charged pions
and require the reduced $\chi^2$ to be less than 10.
The efficiency of this requirement is 94\%.
We require the reconstructed mass of each $\eta$  
candidate to be in the range [500,600] MeV/$c^2$ and 
refit momenta of its daughters with the constraint of the nominal $\eta$ mass.

\begin{figure}[htbp]
\includegraphics[width=0.50\textwidth]{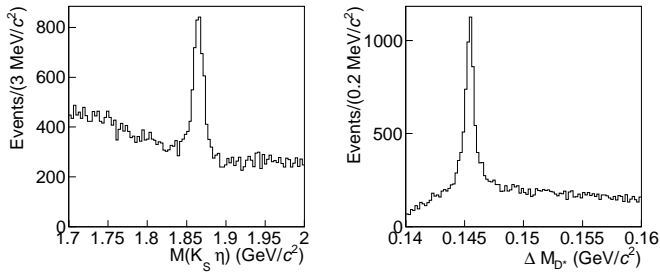}
\caption{
Invariant mass of the $K^0_S\eta$ combinations (left)
and the $D^*$--$D^0$ mass difference (right) 
for $\eta \to \pi^+ \pi^- \gamma$ decays.
}
\label{fig:data_mass}
\end{figure}

\begin{figure}[htbp]
\mbox{
\includegraphics[width=0.40\textwidth]{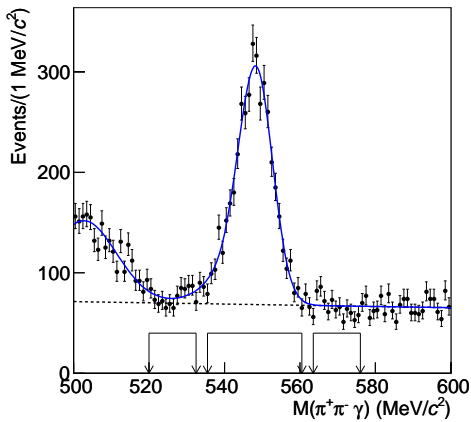}
}
\caption{
Invariant mass distribution of the $\pi^+\pi^-\gamma$ combinations (points
with error bars), fit result (solid curve) and combinatorial
background component (dashed line) 
of the fit function. Arrows with lines
indicate boundaries of the signal and sideband regions.
}
\label{fig:data_eta_B_gauss}
\end{figure}

Combinations of a $K^0_S$ candidate and $\eta$ candidate
are fit to a common vertex and
their invariant mass is required to be within $\pm40$~MeV/$c^2$
of the nominal $D^0$ mass. The $D^0$ and $\pi^+$ combinations 
are fitted to the IP, and the mass difference $\Delta
M_{D^*}=M(K^0_S\eta\pi^+)-M(K^0_S\eta)$ is required to satisfy 
$\Delta M_{D^*} \in [143,148]$~MeV/$c^2$. To remove the combinatorial background, the
momentum of the $D^{*+}$ candidates, measured in the center-of-mass
system, is required to be greater than
2.5, 2.6, and 3.0 GeV/$c$ for the data taken below,
at, and above the $\Upsilon(4S)$
resonance, respectively. 
Figure~\ref{fig:data_mass} shows the invariant mass of the $K^0_S\eta$
combinations (left) and the mass difference (right) 
for $\eta \to \pi^+ \pi^- \gamma$ decays
after applying all selection criteria described above, except the mass
requirements themselves. 
Figure~\ref{fig:data_eta_B_gauss}  shows the invariant mass of the $\pi^+\pi^-\gamma$
combinations after all requirements. There are clear peaks of signal events in all
distributions; the increase of the background at low masses in the
$M(\pi^+\pi^-\gamma)$ distribution is due to the feed-down from the
$\eta\to\pi^+\pi^-\pi^0$ decays 
when a photon from $\pi^0$ is not
reconstructed.

%
%

%
%

To extract the signal yield, we perform a binned 
maximum likelihood
fit to the $M(\pi^+\pi^-\gamma)$
distribution. The fit function is  
the sum of the signal, the combinatorial background
and the feed-down background components.
The signal probability density function
(PDF) is the sum of a Gaussian and a bifurcated Gaussian with the 
ratios of widths fixed from the MC simulation.
A linear function is used for 
the combinatorial background PDF. 
The feed-down contribution is described by a Gaussian with shape parameters
fixed from the MC simulation. The confidence level 
(p-value) of the fit is 12\%
and the $\eta\to\pi^+\pi^-\gamma$ signal yield is 
$N_{\eta} = 2974 \pm 90$ events.
The feed-down yield agrees well with the expectation.

As a cross-check, we measure the ratio of branching fractions
${\mathcal{B}(\eta \rightarrow \pi^+\pi^- \gamma)}/
{\mathcal{B}(\eta \rightarrow \pi^+\pi^- \pi^0)}$.
The fit to the $\pi^+\pi^-\pi^0$ invariant mass
distribution is similar to the one described above, except that the
combinatorial background is described by a second-order polynomial
and there is no feed-down background. The reconstruction efficiencies,
determined from the MC simulation, are 
$\varepsilon (\pi^+\pi^-\gamma) = $ 5.1\% 
and
$\varepsilon (\pi^+\pi^-\pi^0) = $  4.8\%.
The
measured ratio of branching fractions, $0.185\pm0.007$, where the
uncertainty is statistical only, is in
good agreement with the world-average value of
$0.184\pm0.004$~\cite{PDG}.

\begin{figure}[htbp]
\mbox{
\includegraphics[width=0.40\textwidth]{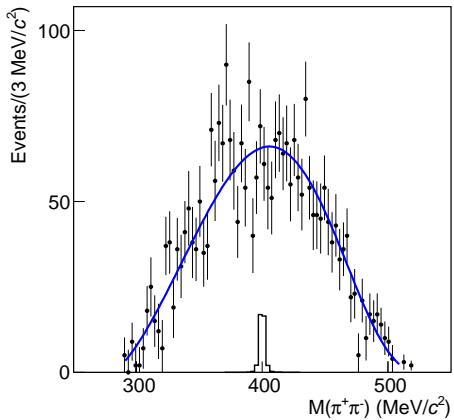}
}
\caption{
$\pi^+ \pi^-$ invariant mass distribution from the $\eta \to \pi^+ \pi^- \gamma$ 
signal (points with error bars), the fitted differential
decay rate described in Eq.~(\ref{eq:dgdx}) (solid curve), and an 
example 
$U^\prime$ signal at a mass of 400 MeV/$c^2$ from 
$\eta \to U^\prime \gamma, U^\prime \to \pi^+ \pi^-$ 
(histogram with arbitrary normalization).
}
\label{fig:data_B}
\end{figure}

We define the $\eta$ signal region as
$M(\pi^+\pi^-\gamma) \in [535.5,560.5]$~MeV/$c^2$, and the sideband regions
used for background subtraction
as $M(\pi^+\pi^-\gamma) \in [520.0,532.5]$ or 
$[563.5,576.0]$~MeV/$c^2$. The $M(\pi^+\pi^-)$
distribution for the background-subtracted $\eta$ signal
is shown in Fig.~\ref{fig:data_B}. 
 
To describe the $M(\pi^+\pi^-)$ distribution, we use an expression of the
differential decay rate based on low-energy
quantum chromodynamics (QCD) phenomenology~\cite{ref:stollenwerk, ref:adlarson} 
using a combination of chiral perturbation theory and dispersive analysis,
\begin{eqnarray} 
\frac{d\Gamma}{d s}
\propto
|P(s) F_V(s)|^2 (m_\eta^2 - s)^3 s (1-4 m^2_\pi/s)^{3/2},
\label{eq:dgdx}
\end{eqnarray} 
where $s\equiv M(\pi^+\pi^-)^2$, $P(s)$ is a reaction-specific
perturbative part, and $F_V(s)$ is the pion vector form factor. We use
$|P(s)| = 1+ (1.89\pm 0.64) s$ ~\cite{ref:adlarson}
and $|F_V(s)| = 1+ (2.12\pm0.01) s + (2.13\pm 0.01) s^2 
+ (13.80\pm 0.14) s^3$~\cite{ref:stollenwerk}
($s$ in GeV$^2/c^4$).
The numerical values and the uncertainties of the expansion coefficients of $|P(s)|$ 
and $|F_V(s)|$ are taken from fits 
to data of 
$\eta^{(\prime)} \to \pi^+ \pi^- \gamma$ decays.
We multiply the $d\Gamma/ds$ expression from Eq.~(\ref{eq:dgdx}) by
the reconstruction efficiency. 
The efficiency as a function of $M(\pi^+ \pi^-)$
is approximately flat but drops to zero at the kinematic
limit of $m_\eta$.
The fit results are presented in Fig.~\ref{fig:data_B}.
Equation~(\ref{eq:dgdx}) describes the $M(\pi^+\pi^-)$ distribution well,
and the confidence level of the fit is 95\%.

We add the $U^\prime$ signal to the above fit function and
perform fits while fixing the $U^\prime$ mass at a value between 290 and 520
MeV/$c^2$ in steps of 1 MeV/$c^2$. 
The $U^\prime$ signal is described by the 
sum of two Gaussians.
The signal resolution of the core Gaussian
is about 1~MeV/$c^2$ near the $2 m_\pi$ threshold
and 2~MeV/$c^2$ at the $m_\eta$ kinematic limit.
An example of the $U^\prime$ signal with the mass of 400 MeV/$c^2$ and
arbitrary normalization is shown in Fig.~\ref{fig:data_B}. We do not find a
significant $U^\prime$ signal at any mass value. The typical uncertainty
in the $U^\prime$ yield $N_{U^\prime}$ is $\mathcal{O}(1-10)$ events.
%
%

%
%
We express the baryonic fine structure constant
$\alpha_{U^\prime}$ using the equation for the partial width ratio
$\Gamma(\eta \rightarrow U^\prime \gamma)/\Gamma(\eta \rightarrow 
\gamma \gamma)$
from Ref.~\cite{ref:tulin} as:
\begin{eqnarray}
\alpha_{U^\prime} &=&
\Bigg[
\frac{\alpha}{2} 
\Bigg(
1-\frac{m^2_{U^\prime}}{m^2_\eta}
\Bigg)^{-3}
\Bigg|
\mathcal{F}(m^2_{U^\prime})
\Bigg|^{-2}
\frac{1}{\mathcal{B}(U^\prime \rightarrow \pi^+ \pi^-)}
\Bigg]
\nonumber \\
&\times&
\Bigg[
\frac{\Gamma(\eta \rightarrow \pi^+ \pi^- \gamma)}{\Gamma(\eta \rightarrow \gamma \gamma)}
\Bigg]
\nonumber \\
&\times&
\Bigg[
\frac{\Gamma(\eta \rightarrow U^\prime \gamma \rightarrow \pi^+ \pi^- \gamma)}
{\Gamma(\eta \rightarrow \pi^+ \pi^- \gamma)}
\Bigg],
\label{eq:alpha_B}
\end{eqnarray}
where $\alpha$ is the electromagnetic fine structure constant. 
The first factor in Eq.~(\ref{eq:alpha_B}), 
which is purely theoretical, contains the phase space,
the form factor $\mathcal{F}(m^2_{U^\prime})$, and 
the branching fraction of $U^\prime  \rightarrow \pi^+ \pi^-$ decay. 
The branching fraction is about 
2-4\%, 
as computed from formulae provided in Ref.~\cite{ref:tulin} and references therein.
The second factor is obtained from the latest measurements~\cite{PDG}.
The third factor is determined from the $\eta$ and $U^\prime$ yields 
and reconstruction efficiencies
${(N_{U^\prime}/\varepsilon(\eta \to U^\prime \gamma \to
\pi^+ \pi^- \gamma))}/{(N_{\eta}/\varepsilon(\eta \to \pi^+ \pi^- \gamma))}$.

%
%
To estimate the systematic uncertainties in the
$\eta\to \pi^+\pi^-\gamma$ and
$\eta\to U^\prime \gamma\to\pi^+\pi^-\gamma$ yields, we change the
parameterization of the combinatorial background in 
the $M(\pi^+ \pi^- \gamma)$ fit from a first- to
a second-order polynomial and account for the background
non-linearity while subtracting the sidebands. The change in the
$\eta$ yield is at the 1\% level, while the change in the $U^\prime$
yield is negligible.
The systematic effect due to the uncertainties of the expansion coefficients 
in $|P(s)|$ and $|F_V(s)|$ is negligible in the $U^\prime$ yield.
The systematic uncertainty in the ratio of the reconstruction efficiencies 
$\varepsilon(\eta\to U^\prime \gamma\to\pi^+\pi^-\gamma)/
\varepsilon(\eta\to\pi^+\pi^-\gamma)$ is conservatively estimated to
be 4\% (1\% per track and 3\% per photon).
The total systematic uncertainties are estimated by adding the above contributions
in quadrature.

\begin{figure}[htbp]
\includegraphics[width=0.40\textwidth]{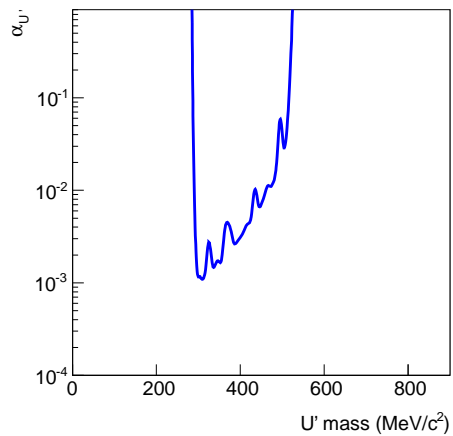}
\caption{
Computed 95\% upper limit on the baryonic fine structure constant
$\alpha_{U^\prime}$ as a function of the unknown $U^\prime$ mass
(solid curve).
}
\label{fig:data_limit}
\end{figure}

Using Eq.~(\ref{eq:alpha_B}),
we set a 95\% confidence level upper limits on $\alpha_{U^\prime}$ 
using the Feldman-Cousins approach~\cite{ref:fc},
adding the statistical and systematic uncertainties in quadrature.
The upper limit 
as a function of the $U^\prime$ boson mass is shown in Fig.~\ref{fig:data_limit}.
Considering other results in this mass region, 
we find that our limit
is stronger than that from a model-dependent analysis~\cite{ref:tulin}
of the $\phi\to e^+e^-\gamma$ decays~\cite{ref:kloe} for
$m_{U^\prime}>450$~MeV/$c^2$, but weaker than the limit based on the
$\eta\to\pi^0\gamma\gamma$ total rate~\cite{ref:tulin}. 
During preparation of this manuscript,
we learned that the data set in Ref.~\cite{ref:babusci} contains many more
$\eta \rightarrow \pi^+ \pi^- \gamma$ decays and can provide a more stringent 
limit on $\alpha_{U^\prime}$ in future.

%
%

To conclude, we perform a search for a dark vector gauge boson
$U^\prime$ that couples to quarks~\cite{ref:tulin}, using the decay chain
$D^{*+} \to D^0 \pi^+, D^0 \to
K^0_S \eta, \eta \to U^\prime \gamma, U^\prime \to \pi^+ \pi^-$.
Our results limit the
baryonic fine structure constant $\alpha_{U^\prime}$ to below
$10^{-3}-10^{-2}$ at 95\% confidence level over the $U^\prime$ mass
range 290 to 520~MeV/$c^2$. This is the first search for $U^\prime$
in the $\pi^+\pi^-$ mode.
We find that our limit
is stronger than that from a model-dependent analysis~\cite{ref:tulin}
of the $\phi\to e^+e^-\gamma$ decays~\cite{ref:kloe} for
$m_{U^\prime}>450$~MeV/$c^2$, but weaker than the limit based on the
$\eta\to\pi^0\gamma\gamma$ total rate~\cite{ref:tulin}. \\

We thank the KEKB group for the excellent operation of the
accelerator; the KEK cryogenics group for the efficient
operation of the solenoid; and the KEK computer group,
the National Institute of Informatics, and the 
PNNL/EMSL computing group for valuable computing
and SINET4 network support.  We acknowledge support from
the Ministry of Education, Culture, Sports, Science, and
Technology (MEXT) of Japan, the Japan Society for the 
Promotion of Science (JSPS), and the Tau-Lepton Physics 
Research Center of Nagoya University; 
the Australian Research Council;
Austrian Science Fund under Grant No.~P 22742-N16 and P 26794-N20;
the National Natural Science Foundation of China under Contracts 
No.~10575109, No.~10775142, No.~10875115, No.~11175187, No.~11475187
and No.~11575017;
the Chinese Academy of Science Center for Excellence in Particle Physics; 
the Ministry of Education, Youth and Sports of the Czech
Republic under Contract No.~LG14034;
the Carl Zeiss Foundation, the Deutsche Forschungsgemeinschaft, the
Excellence Cluster Universe, and the VolkswagenStiftung;
the Department of Science and Technology of India; 
the Istituto Nazionale di Fisica Nucleare of Italy; 
the WCU program of the Ministry of Education, National Research Foundation (NRF) 
of Korea Grants No.~2011-0029457,  No.~2012-0008143,  
No.~2012R1A1A2008330, No.~2013R1A1A3007772, No.~2014R1A2A2A01005286, 
No.~2014R1A2A2A01002734, No.~2015R1A2A2A01003280 , No. 2015H1A2A1033649;
the Basic Research Lab program under NRF Grant No.~KRF-2011-0020333,
Center for Korean J-PARC Users, No.~NRF-2013K1A3A7A06056592; 
the Brain Korea 21-Plus program and Radiation Science Research Institute;
the Polish Ministry of Science and Higher Education and 
the National Science Center;
the Ministry of Education and Science of the Russian Federation and
the Russian Foundation for Basic Research;
the Slovenian Research Agency;
Ikerbasque, Basque Foundation for Science and
the Euskal Herriko Unibertsitatea (UPV/EHU) under program UFI 11/55 (Spain);
the Swiss National Science Foundation; 
the Ministry of Education and the Ministry of Science and Technology of Taiwan;
and the U.S.\ Department of Energy and the National Science Foundation.
This work is supported by a Grant-in-Aid from MEXT for 
Science Research in a Priority Area (``New Development of 
Flavor Physics'') and from JSPS for Creative Scientific 
Research (``Evolution of Tau-lepton Physics'').
EW acknowledges partially NRF grant of Korea Grants No. NRF-2011-0030865 
and Korea University Future Research Grant,
and thanks B. R. Ko for his suggestion on an inclusive version 
of this analysis.




\begin{thebibliography}{99}
\bibitem{ref:sign}
Throughout this paper, inclusion of the charge-conjugate decay is implied 
unless stated otherwise.
\end{thebibliography}
\end{document}